\documentclass[preprint,aps]{revtex4}

\usepackage{graphicx}
\usepackage{dcolumn}

\begin{document}

\preprint{Lebed-JETP-Letters}

\title{Type-IV Superconductivity: Cooper Pairs 
with Broken Inversion  and Time-Reversal 
Symmetries in Conventional Superconductors}

\author{A.G. Lebed$^*$}

 \affiliation{Department of Physics, 
University of Arizona,
1118 E. 4-th Street, Tucson, AZ 
85721, USA}

\begin{abstract}
Vortex phase in a singlet superconductor in the 
absence of impurities is shown to be  absolutely unstable 
with respect to the appearance of a triplet component which 
breaks both inversion and  time-reversal symmetries of Cooper 
pairs.
Symmetry breaking paramagnetic effects are demonstrated
to be of the order of unity if the orbital upper critical field, $H_{c2}(0)$,
is of the order of  Clogston paramagnetic limiting field, 
$H_p$.
We suggest a generic phase diagram of such type-IV
superconductor, which is singlet one at $H=0$ and characterized 
by mixed singlet-triplet order parameter with broken time-reversal 
symmetry in vortex 
phase.
A possibility to observe type-IV superconductivity in clean
organic, high-T$_c$, MgB$_2$, and  other superconductors
is discussed.
 \\ \\ PACS numbers: 74.20.Rp, 74.25.Op, 74.70.-b
\end{abstract}

\maketitle
  
\pagebreak
  
It is well known that Meissner effect, which is the main 
feature of superconductivity phenomenon, is used to classify
superconducting materials.
In type-I superconductors, where Meissner effect is complete, 
superconductivity is destroyed at $H > H_c$ (where $H_c$ is thermodynamic 
critical field), whereas, in type-II superconductors, superconductivity 
phenomenon survives at highier magnetic  fields, $H_{c1} < H < H_{c2}$ 
(where $H_{c1} < H_c < H_{c2}$)  in the form of Abrikosov vortices 
[1,2].
Type-II superconductors can be subdivided into two main classes:
superconducting alloys (or dirty superconductors) [1,2] and relatively 
clean superconductors, where type-II superconductivity is due to anisotropy 
of their electron spectra and relatively high effective masses of quasi-particles 
[3].
Due to a success in synthesis of novel materials, a number of new 
classes of relatively clean type-II superconductors were discovered 
during last 30 years, including organic [4,5], heavy-fermion [6], 
high-T$_c$ [7], Sr$_2$RuO$_4$ [8], MgB$_2$ [9], and other 
superconductors.
Currently, the above mentioned relatively clean type-II superconductors are 
the most interesting and important materials both from fundamental 
point of view and from point of view of their possible 
applications.

Usually, orbital superconducting order parameter, $\Delta({\bf r_1},{\bf r_2})$, 
corresponding to pairing of two electrons in Cooper pair, can  be expressed 
in the form 
$\Delta({\bf r}_1, {\bf r}_2) =  \Delta({ \bf R}) \hat{ \Delta} ({\bf r})$.
[Here, external order parameter, $\Delta({\bf R})$, is related to  motion 
of a center of mass of Cooper pair, ${\bf R}=({\bf r_1}+{\bf r_2})/2$, 
whereas internal order parameter, $\hat{\Delta } ({\bf r})$, describes 
relative motion of electrons in Cooper pair,
${\bf r}={\bf r_1}-{\bf r_2}$].
From this point of view, type-II superconductors in their vortex 
phases are characterized by broken symmetries of external 
order parameter, $\Delta({ \bf R} )$, which is responsible for Meissner 
currents [1,2].

Other important issues are symmetries of internal orbital order parameter, 
$\hat{ \Delta} ({\bf r})$ (or its Fourier component $\hat{ \Delta} ({\bf k})$), 
and related spin part of superconducting order 
parameter, $\hat {\Delta} (\sigma_1, \sigma_2)$.
In accordance with Fermi statistics, internal order parameter, 
$\hat{ \Delta} ({\bf k})$, is an even function of variable ${\bf k}$
in the case of singlet superconductivity (where the total spin of Cooper 
pair $|{\bf S}|= 0$) , whereas $\hat{ \Delta} ({\bf k})$ is an odd function 
of ${\bf k}$ in the case of triplet superconductivity (where $|{\bf S}| = 1$)
[10,11].
Depending on symmetry properties of $\hat{ \Delta} ({\bf k})$, 
superconductors are subdivided into conventional ones [1,2]
(where superconductivity can be described in terms of BCS s-wave 
singlet pairing) and unconventional ones [10,11] (where symmetry of 
$\hat{ \Delta} ({\bf k})$ is lower than the underlying symmetry of 
crystalline lattice).

It is commonly believed [1,2,10,11] that magnetic field does not change 
internal superconducting order parameters (i.e., wave functions 
$\hat{ \Delta} ({\bf k})$ and $\hat{\Delta}(\sigma_1, \sigma_2)$) 
and, thus, Cooper pairs can be considered as unchanged elementary 
particles in Abrikosov 
vortex phase. 
Moreover, although related to the external degrees of freedom Meissner 
currents break time-reversal symmetry of $\Delta({\bf R})$, internal orbital 
and spin order parameters, $\hat{ \Delta} ({\bf k})$ and 
$\hat{\Delta}(\sigma_1, \sigma_2)$, 
are believed to preserve $t \rightarrow -t$
symmetry. 
The main goal of our Letter is to show that there have to exist
type-IV superconductors [12], which exhibit qualitatively different
magnetic properties and are characterized by Cooper pairing with 
broken time reversal, $t \rightarrow -t$, and inversion, 
${\bf k} \rightarrow {\bf -k}$, symmetries of  internal  order parameters
in vortex
phase.
[Note that we define type-IV superconductivity as singlet superconductivity 
at $H=0$ and in Meissner phase which exhibits  broken symmetries of Cooper 
pairs internal wave functions 
in vortex phase.]

More precisely, below we suggest and prove the following theorem: 
each singlet type-II superconductor in the absence of impurities is 
actually type-IV superconductor with broken $t \rightarrow -t$ and 
${\bf k} \rightarrow {\bf -k}$ symmetries of internal order parameters 
in the vortex phase, provided that effective constant, responsible 
for triplet (p-wave) superconducting pairing, is not  
zero, $g_t \neq 0$.
We show that the above mentioned theorem is due to careful account 
for paramagnetic spin splitting effects in vortex phase, which 
have been treated so far only in the case $g_t=0$ 
[1,2,10,11,13].
In particular, we demonstrate that superconducting internal order 
parameter is a mixture of a singlet component, $ \hat{ \Delta_s} ({\bf k})$,
with a triplet component, $i \ \hat{ \Delta_t} ({\bf k})$, which breaks both 
inversion, ${\bf k} \rightarrow -{\bf k}$, and time-reversal,
$t \rightarrow -t$, symmetries due to an imaginary 
coefficient $i$.
We point out that the above mentioned effects of singlet-triplet mixing 
are expected to be of the order of unity in a number of modern relatively 
clean type-II superconductors, where $H_{c2}(0)  \sim H_p$ (see discussion
in the end of the Letter). 
[Here, $H_{c2}(0)$ is orbital upper critical field at $T=0$ [1,2,10,11] 
and $H_p$ is  Clogston paramagnetic
 limiting field [13,1].

Although, in the Letter, we consider vortex phase with broken 
symmetries only in Ginzburg-Landau (GL) region of a s-wave 
layered superconductor in a parallel magnetic field, we stress 
that the suggested theorem is very general and based only on 
symmetry
arguments. 
As we argue below, the above mentioned theorem is a consequence 
of broken spin symmetry  (due to paramagnetic spin splitting effects) [13] 
and broken translational symmetry of external orbital order parameter, 
$\Delta({\bf R})$,
in vortex phase.
As a result, the theorem, suggested in the Letter, has to be valid for any 
s- and d-wave singlet superconductor [14] for both attractive  and repulsive 
electron-electron interactions in a triplet (p-wave) 
channel. 

In other words, the main our message is that Cooper pairs 
cannot be considered as unchanged elementary particles in 
a magnetic field in modern type-II superconductors, 
where 
$H_{c2} (0) \sim H_p$.
As shown below, magnetic fields of the order of $H \sim H_p$  
qualitatively change the nature of Cooper pairs in 
vortex phase.
We suggest that, in relatively clean conventional type-II superconductors, 
there have to exist the forth critical magnetic field, $H_{c4}(T)$, corresponding 
to phase transition (or crossover) between Abrikosov vortex phase and exotic 
vortex phase, where inversion and time-reversal symmetries of Cooper pairs 
are broken and, thus, topologic properties of vortices are
unusual (see Fig.1). 

At first, let us qualitatively explain why paramagnetic effects
lead to the appearance of a triplet component in vortex phase 
of a  conventional singlet 
superconductor.
It is well known [10,11] that spin component 
of a singlet order parameter is antisymmetric function of spin
variables, ${\hat \Delta}_s(+, -) = 
- {\hat \Delta}_s(-, +)$.
In the presence of Abrikosov vortices, external order parameter,
$\Delta({\bf R})$, varies with coordinate ${\bf R}$, which corresponds
to superconducting pairing of electrons with non-zero total 
momenta of Cooper pairs of the order of $|{\bf p}| \simeq \hbar / \xi $,
where $\xi$ is a coherence length [1,2,10,11].
Let us consider superconducting pairing of two electrons with
total momentum ${|{\bf p_0}| \neq 0}$ in the presence of spin splitting
paramagnetic effects (see Fig.2).
As it is seen from Fig.2, absolute value of spin component
$\Delta(+, -)$ is not equal to absolute value of spin component
 $\Delta(-, +)$ if ${|{\bf p_0}| \neq 0}$.
 This means that $\Delta(+, -) \neq - \Delta(-, +)$ and, thus, 
 in addition to singlet order parameter, a triplet component, 
 corresponding to superconducting pairing with $|{\bf S}| =1$ and $S_z=0$,
  appears, where $S_z$ is a component of total spin of Cooper 
  pair along magnetic 
 field direction.
 
 Below, we quantitatively  describe superconducting pairing with
 internal order parameter, exhibiting broken inversion and
 time-reversal symmetries, in a singlet s-wave superconductor
 with layered electron spectrum,  
\begin{equation}
\epsilon_0({\bf p})= (p^2_x + p^2_y) / 2m + 2 t_z \cos(p_z d) \ , \ \ \ 
\epsilon_F = m v^2_F / 2 \ ,
\end{equation}
in a magnetic field:
\begin{equation}
{\bf H} = (0,H,0) \  ,  \ \ \ \ {\bf A} = (0,0,-Hx) \ .
\end{equation}
In case, where electron-electron interactions do not 
depend on electron spins, the total Hamiltonian of electron 
system can be written as follows:
\begin{eqnarray}
 &&\hat{H} = \hat{H}_0 + \hat{H}_{int} \ , \ \ \ \ \ 
\hat{H}_0 = \sum_{{\bf p} ,  \sigma}  
\epsilon_{\sigma} ({\bf p})  \ c^{+}_{\sigma} ({\bf p}) \ c_{\sigma} ({\bf p}) \ ,
\nonumber\\ 
&&\hat{H}_{int} = \frac{1}{2} \sum_{{\bf q} ,  \sigma}  \sum_{{\bf p} , {\bf p_1}}  \ 
U({\bf p} , {\bf p_1}) \ 
c^{+}_{\sigma} ({\bf p} + \frac{{\bf q}}{2})  \ c^{+}_{-\sigma} ({-\bf p} + \frac{{\bf q}} {2})
\ c_{-\sigma} ({-\bf p_1} + \frac{{\bf q}}{2}) \  c_{\sigma} ({\bf p_1} + \frac{{\bf q}}{2}) \ ,
\end{eqnarray} 
where $\sigma = \pm 1$, $\epsilon_{\sigma} ({\bf p}) = \epsilon_0({\bf p}) - 
\sigma \mu_B H$, $c^{+}_{\sigma} ({\bf p})$ and $c_{\sigma} ({\bf p})$
are electron creation and annihilation operators. 
As usually [10,11], electron-electron interactions are subdivided 
into singlet and triplet channels:
\begin{eqnarray}
&&U({\bf p}, {\bf p_1}) = U_s({\bf p}, {\bf p_1}) + U_t({\bf p}, {\bf p_1}) \ , \ \ 
U_s({\bf p}, {\bf p_1}) = U_s(-{\bf p}, {\bf p_1}) = U_s({\bf p}, - {\bf p_1}) \ ,
\nonumber\\ 
&&U_t({\bf p}, {\bf p_1}) = - U_t(-{\bf p}, {\bf p_1}) = - U_t({\bf p}, - {\bf p_1})    \ .
\end{eqnarray} 
Below, we define normal and anomalous (Gorkov) Green functions 
by standard way [15,16]:
\begin{eqnarray}
&&G_{\sigma, \sigma}({\bf p}, {\bf p_1}; \tau)  = 
- < T_{\tau} c_{\sigma}({\bf p}, \tau) c^+_{\sigma}({\bf p_1}, 0 > , \ \  
F_{\sigma, -\sigma}({\bf p}, {\bf p_1}; \tau)  = 
< T_{\tau} c_{\sigma}({\bf p}, \tau) c_{-\sigma}({-\bf p_1}, 0) > \ ,
\nonumber\\ 
&&F^+_{\sigma, -\sigma}({\bf p}, {\bf p_1}; \tau)  = 
< T_{\tau} c^+_{\sigma}({-\bf p}, \tau) c^+_{-\sigma}({\bf p_1}, 0)  >   \ ,
\end{eqnarray} 
where $<...>$ stands for Gibbs averaging procedure with
Hamiltonian (3). $\\$

If we define singlet and triplet superconducting order parameters [10,11],
\begin{eqnarray}
&&\Delta_s ({\bf p}, {\bf q}) = - \frac{1}{2} \sum_{ {\bf p}_1}  U_s({\bf p},{\bf p_1}) 
T \sum_{\omega_n} 
[F_{+, -}(i \omega_n ; {\bf p}_1+ \frac{{\bf q}}{2}, {\bf p_1}- \frac{{\bf q}}{2}) 
- F_{-, +}(i \omega_n ; {\bf p}_1+\frac{{\bf q}}{2}, {\bf p_1}-\frac{{\bf q}}{2})]  \ , 
\nonumber\\
&&\Delta_t ({\bf p}, {\bf q}) = - \frac{1}{2} \sum_{ {\bf p}_1}  U_t({\bf p},{\bf p_1}) 
T \sum_{\omega_n} 
[F_{+, -}(i \omega_n ; {\bf p}_1+\frac{{\bf q}}{2}, {\bf p_1}-\frac{{\bf q}}{2})
+ F_{-, +}(i \omega_n ; {\bf p}_1+\frac{{\bf q}}{2}, {\bf p_1}-\frac{{\bf q}}{2})]  \ , 
\end{eqnarray} 
then, using Green-function technique [10,11,15,16], we obtain 
the following equations:
\begin{eqnarray}
&&[i \omega_n - \epsilon_{\sigma}({\bf p})]  G_{\sigma, \sigma}(i \omega_n; {\bf p}, {\bf p_1}) 
+ \sum_{\bf q} [\sigma \Delta_s ({\bf p}, {\bf q} )  + \Delta_t ({\bf p}, {\bf q} )] 
F^+_{-\sigma, \sigma} (i \omega_n; {\bf p}-{\bf q}, {\bf p_1})= \delta ({\bf p} - {\bf p_1}) \ , 
\nonumber\\ 
&&[i \omega_n - \epsilon_{\sigma}({\bf p})]  F_{\sigma, -\sigma}(i \omega_n; {\bf p}, {\bf p_1}) 
- \sum_{\bf q} [\sigma \Delta_s ({\bf p}, {\bf q} )  + \Delta_t ({\bf p}, {\bf q} )] 
G_{-\sigma, -\sigma} (- i \omega_n; -{\bf p_1}, -{\bf p}+{\bf q})= 0    \  ,
\nonumber\\ 
&&[i \omega_n + \epsilon_{\sigma}({\bf p})]  F^+_{\sigma, -\sigma}(i \omega_n; {\bf p}, {\bf p_1}) 
+ \sum_{\bf q} [\sigma \Delta^+_s ({\bf p}, {\bf q} )  + \Delta^+_t ({\bf p}, {\bf q} )] 
G_{\sigma, \sigma} ( i \omega_n; {\bf p} + {\bf q}, {\bf p_1})= 0    \  ,
\end{eqnarray} 
which extend Gorkov equations [15,16] to case of 
coexistence of singlet and triplet order parameters (6).
[Note that Eqs.(7), suggested in the Letter, are rather general 
and describe coexistence of triplet and singlet order
parameters  for spin independent electron-electron
interactions at arbitrary temperatures].

The goal of our Letter is to solve Eqs.(7) in the case, where  layered 
superconductor  (1) is placed in a parallel magnetic field (2).
Below, we  consider a phase transition line between metallic and 
singlet-triplet mixed superconducting phases in GL
region (i.e., at $(T_c-T)/ T_c \ll 1$), where $T_c$ is a 
transition temperature from metallic state to s-wave singlet  
phase at $H=0$.
For this purpose, we linearize [17,10,11] Eqs.(7) with respect to superconducting 
order parameters, $\Delta_s({\bf p}, {\bf q})$ and 
$\Delta_t({\bf p}, {\bf q})$.
As a result, we obtain the following system of linear equations:
\begin{eqnarray}
&&\Delta_s ({\bf p}, {\bf q}) = - \frac{1}{2} \sum_{ {\bf p}_1}  U_s ({\bf p},{\bf p_1}) 
T \sum_{\omega_n} [ \Delta_s ({\bf p_1}, {\bf q}) S_{+}(i \omega_n; {\bf p_1}, {\bf q}) 
+ \Delta_t ({\bf p_1}, {\bf q}) S_{-}(i \omega_n; {\bf p_1}, {\bf q})]  \ , 
\nonumber\\
&&\Delta_t ({\bf p}, {\bf q}) = - \frac{1}{2} \sum_{ {\bf p}_1}  U_t ({\bf p},{\bf p_1}) 
T \sum_{\omega_n} [ \Delta_t ({\bf p_1}, {\bf q}) S_{+}(i \omega_n; {\bf p_1}, {\bf q}) 
+ \Delta_s ({\bf p_1}, {\bf q}) S_{-}(i \omega_n; {\bf p_1}, {\bf q})]   \ , 
\nonumber\\
&&S_{+}(i \omega_n; {\bf p_1}, {\bf q})  =
G^0_{+} (i \omega_n, {\bf p}_1+ \frac{{\bf q}}{2})
G^0_{-} (-i \omega_n, -{\bf p}_1+ \frac{{\bf q}}{2}) +
G^0_{-} (i \omega_n, {\bf p}_1+ \frac{{\bf q}}{2})
G^0_{+} (-i \omega_n, -{\bf p}_1+ \frac{{\bf q}}{2})   \ , 
\nonumber\\
&&S_{-}(i \omega_n; {\bf p_1}, {\bf q})]  =   
G^0_{+} (i \omega_n, {\bf p}_1+ \frac{{\bf q}}{2})
G^0_{-} (-i \omega_n, -{\bf p}_1+ \frac{{\bf q}}{2}) -
G^0_{-} (i \omega_n, {\bf p}_1+ \frac{{\bf q}}{2})
G^0_{+} (-i \omega_n, -{\bf p}_1+ \frac{{\bf q}}{2}) \ ,
\end{eqnarray} 
where $G^0_{\sigma} (i \omega_n, {\bf p}) = 
1 / [i \omega_n - \epsilon_{\sigma}({\bf p})]$ 
is Green function of a free electron in the presence
of paramagnetic spin splitting effects.
One of the main results of our Letter is that terms
with $S_{-}(i \omega_n, {\bf p}_1, {\bf q})$, mixing
singlet and triplet superconducting pairings, are not 
equal to zero
(see Fig.2).
Therefore, in Abrikosov vortex phase, singlet 
component of superconducting order parameter
always coexists with triplet one which breaks 
inversion symmetry of Cooper 
pairs.

As an example, let us consider coexistence of singlet 
s-wave and triplet p-wave order parameters, where 
\begin{equation}
U_s({\bf p},{\bf p}_1)  = - (2 \pi /v_F) \  g_s \  ,  \ \ 
U_t({\bf p},{\bf p}_1)  = - (4 \pi/v_F)\  g_t \ \cos (\phi - \phi_1) \ , \ \  g_s > 0, 
\ \ g_s > g_t \ ,
\end{equation}
with $\phi$ and $\phi_1$ being polar angles corresponding to 
momenta ${\bf p}$ and ${\bf p_1}$, respectively.
[Note that inequalities $g_s > 0$ and $g_s > g_t$ correspond
to stabilization of a singlet $s$-wave superconductivity 
at $H=0$].
After substituting Eqs.(9) in Eqs.(8), we can represent singlet
and triplet components of the order parameter as follows: 
$\Delta_s({\bf p}, {\bf q}) = \Delta_s({\bf q})$ and
$\Delta_t({\bf p}, {\bf q}) = \cos \phi \ \Delta_t({\bf q})$,
which satisfy the equations:
\begin{equation}
\Delta_s({\bf q}) = g_s \ A \ \Delta_s({\bf q}) + g_s \ B \ \Delta_t({\bf q}) \ ,  \ \ \ 
\Delta_t({\bf q}) = g_t \ C \ \Delta_t({\bf q}) + g_t \ D \ \Delta_s({\bf q}) \ .
\end{equation} 
Since z-component of vector potential (2) depends only on 
coordinate $x$, we may consider  $q_y=0$ in 
Eqs.(10).
Below,  we calculate quantities $A$, $B$, $C$, and $D$ in 
GL region [3,18,10,11] which corresponds to their expansions 
as power series in small parameters
$v_F q_x /T_c \ll 1$, $t_z d q_z /T_c \ll 1$, and
$\mu_B H /T_c \ll 1$.
As a result, we obtain:
\begin{eqnarray}
&&A  \simeq (\pi T) \sum^{\Omega}_{\omega_n >0} 
\biggl[ \frac{2}{\omega_n} - \frac{1}{4 \omega^3_n} (v^2_Fq^2_x + 
4 t^2_z q^2_z d^2 + 8 \mu^2_b H^2) \biggl]
\ ,
B \simeq - \sqrt{2} \mu_B H v_F q_x (\pi T_c) \sum^{\infty}_{\omega_n >0} \frac{1}{\omega^3_n}
   \ , 
\nonumber\\
&&C \simeq (\pi T) \sum^{\Omega}_{\omega_n >0} 
\biggl[ \frac{2}{\omega_n} - \frac{1}{4 \omega^3_n} (3 v^2_Fq^2_x /2 + 
4 t^2_z q^2_z d^2 + 8 \mu^2_b H^2) \biggl]
 \ , \ 
D = B \ ,
\end{eqnarray} 
where $\Omega$ is a cut-off energy.

We introduce magnetic field (2) in Eqs.(11) using a standard 
quasi-classical eiconal approximation [17,18,10,11]:
$ q_x \rightarrow - i ( d / dx) , \ q_z / 2 \rightarrow eA_z / c = e H x / c$,
which leads to the following GL equations extended to the case of 
triplet-singlet coexistence:

\begin{eqnarray}
&&\biggl[ \tau + \xi^2_{\parallel} \frac{d^2}{dx^2} - \frac{(2\pi \xi_{\perp})^2}{\phi^2_0} H^2 x^2
\biggl] \Delta_s(x) + i \frac{ \sqrt{7 \zeta (3)} }{\sqrt{2} \gamma} \xi_{\parallel}
\biggl( \frac{H}{H_p} \biggl) \frac{d \Delta_t (x)}{dx}=0
   \ , 
 \nonumber\\
&&\biggl[ \frac{g_t - g_s}{g_t g_s} + \frac{3}{2} \xi^2_{\parallel} \frac{d^2}{dx^2} - 
\frac{(2\pi \xi_{\perp})^2}{\phi^2_0} H^2 x^2
\biggl] \Delta_t(x)+
 i \frac{ \sqrt{7 \zeta (3)} }{\sqrt{2} \gamma} \xi_{\parallel}
\biggl( \frac{H}{H_p} \biggl) \frac{d \Delta_s (x)}{dx}=0 \ ,
\end{eqnarray} 
where $\tau = (T_c -T)/T_c \ll 1$, $\xi_{\parallel} = \sqrt{7 \zeta (3)} v_F / 
4 \sqrt{2} \pi T_c$, $\xi_{\perp} = \sqrt{7 \zeta (3)} t_z d / 
2 \sqrt{2} \pi T_c$ are GL coherence lengths [3,1,2], $\phi_0$ is a flux
quantum, $\zeta(z)$ is Riemann zeta-function, $\gamma$ is Euler constant,
$H_p$ is Clogston paramagnetic limiting field [13].

In case, where $g_s - g_t \sim g_s$, Eqs. (12) have the following
solutions:

\begin{equation}
\Delta_s(x) = \exp \biggl(- \frac{\tau x^2}{ 2 \xi^2_{\parallel}} \biggl)   \ , 
\ \ \ \Delta_t(x) =  i \frac{ \sqrt{7 \zeta (3)} }{ \gamma} 
\biggl(  \frac{ g_t g_s }{ g_t - g_s }  \biggl)
\sqrt{\tau} \biggl( \frac{H}{H_p} \biggl) 
\biggl( \frac{ \sqrt{\tau} x }{\sqrt{2} \xi_{\parallel}} \biggl)
 \exp \biggl(- \frac{\tau x^2}{ 2 \xi^2_{\parallel}} \biggl)  .
\end{equation} 
Eqs.(12),(13) are the main results of our Letter. 
They extend GL equations [3,1,2] and their famous Abrikosov 
solution for superconducting nucleus [19,1,2] to the case
$g_t \neq 0$. 
Eqs.(12),(13) directly demonstrate that Abrikosov 
solution [19,1,2] is absolutely unstable in the absence of impurities 
and, thus, singlet order parameter is always mixed with  
triplet one in vortex phase for both attractive, 
$- g_t < 0$, and repulsive, $-g_t >0$, interactions in triplet (p-wave) 
channel.
From Eqs.(12),(13), it also follows that triplet component breaks
not only inversion symmetry, but also time reversal symmetry
since $\Delta^*_t(x)  \neq \Delta_t(x)$ due to imaginary
coefficient $i$. 

 We hope that our results (8)-(13) open new area
 of research: theoretical and experimental studies of exotic 
 vortex superconducting phases
 in singlet superconductors with their properties being even
 more unusual than that in the so-called unconventional
 superconductors [10,11].
 We stress that type-IV superconductivity phenomenon, suggested
 in the Letter, is inherent and very common property of singlet
 superconductivity. 
 In fact, we have shown that each s-wave [14] pure type-II
 superconductor is actually type-IV superconductor.
 The finite amount of impurities may result in the appearance 
 of the forth critical field, $H_{c4}(T)$, which may correspond
 to phase transition (or crossover) between phase, $H_{c1} (T) 
 < H < H_{c4}(T)$, where broken symmetries of Cooper pairs 
 do not exist (or marginal) and phase, $H_{c4} (T) < H < 
 H_{c2} (T)$, where broken inversion and time-reversal 
 symmetries are essential 
 (see Fig.1). 
 From Eqs.(13), it is directly seen that symmetry breaking triplet
 component is of the order of unity at low temperatures in
 such modern strongly correlated superconductors as organic, 
 high-T$_c$, MgB$_2$, and some others, where $|g_t| \sim |g_s|$
 and $H_{c2}(0) \sim H_p$.
 In conclusion, we point out that spin splitting and 
broken  translational symmetry effects have been studied 
in Refs.[20-23] in different 
context.

The author devote results, obtained in this Letter, to his wife
Natalia, whose enormous support have given him a courage to
set and to attack type-IV superconductivity 
problem.

$^*$ Also Landau Institute for Theoretical Physics, 
2 Kosygina Street, Moscow, Russia.

\pagebreak

\begin{figure}[h]
\includegraphics[width=7.6in,clip]{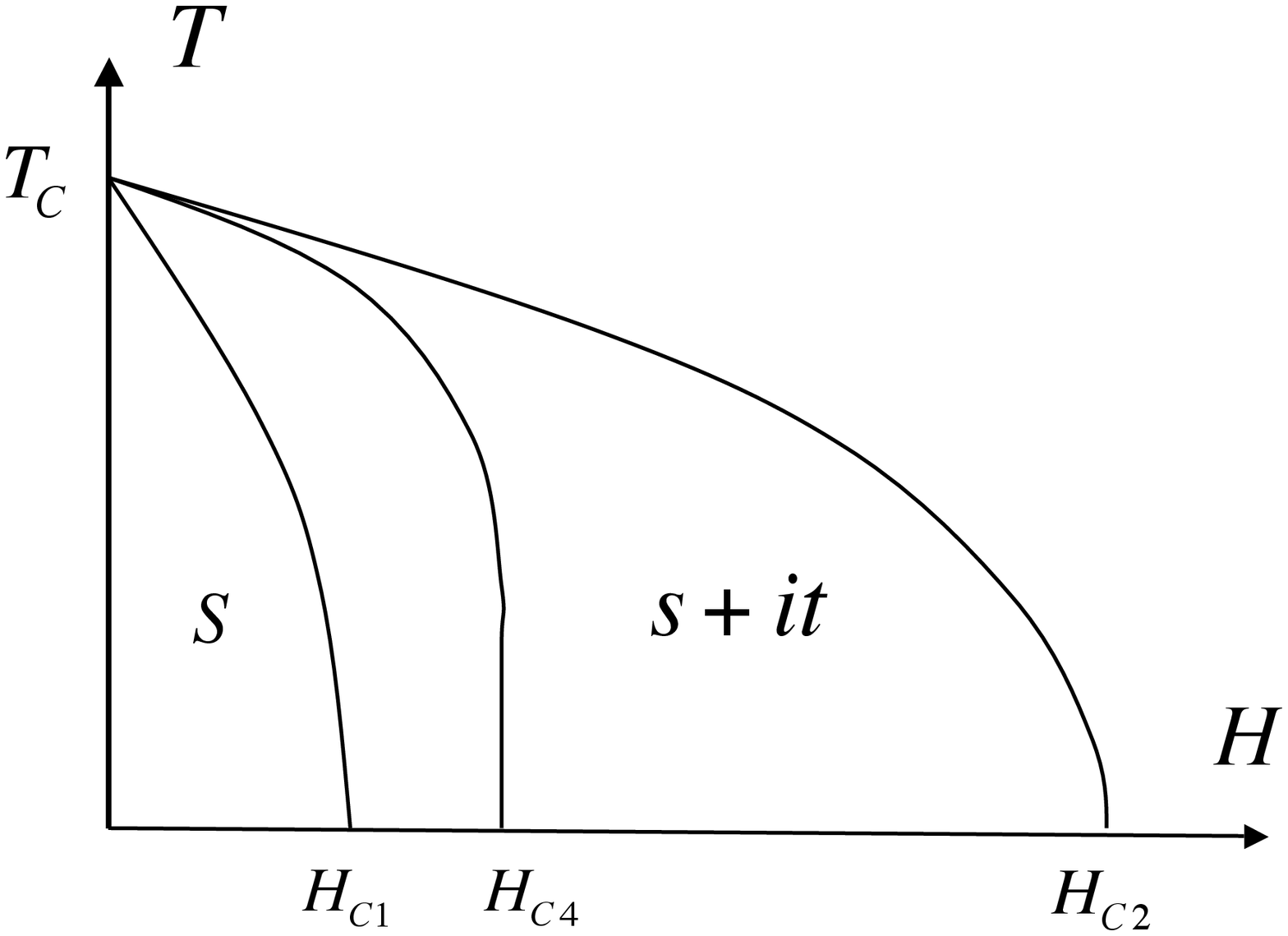}
\caption{ 
A phase diagram of type-IV superconductor,
which is singlet one at $H=0$ and in Meissner
phase, $0 < H < H_{c1}(T)$, and characterized
by broken inversion and time-reversal symmetries
at low temperatures and high magnetic fields,
$H_{c4}(T) < H < H_{c2}(T)$.
In the intermediate region of magnetic fields, 
$H_{c1}(T) < H < H_{c4}(T)$, broken symmetries 
may be absent or marginal.
}
\label{fig1}
\end{figure}

\begin{figure}[h]
\includegraphics[width=7.6in,clip]{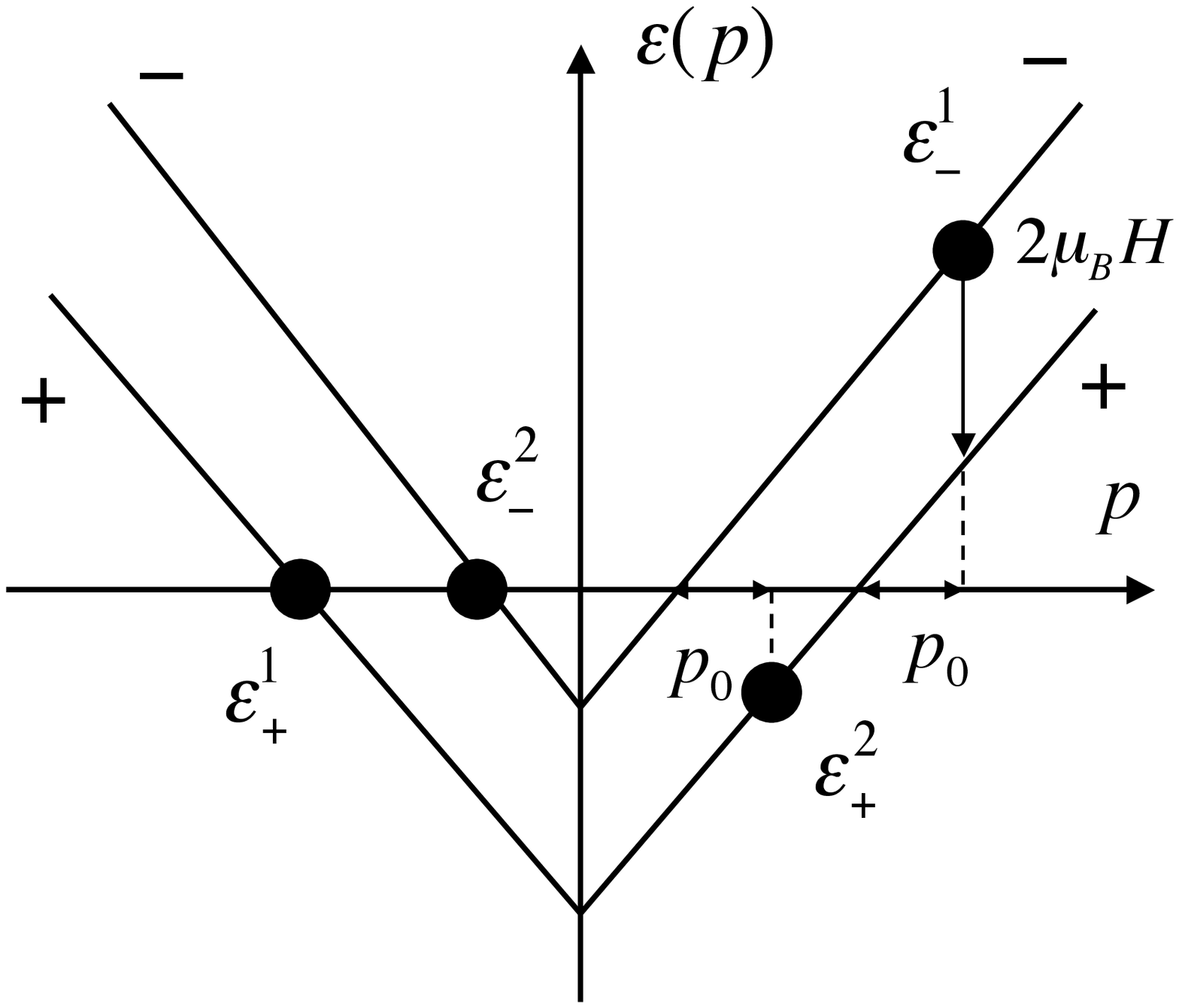}
\caption{ 
In a magnetic field, electron spectra with spin up and spin down 
are split, $\epsilon_{+}(p) = \epsilon_{0}(p) - \mu_B H$ and 
$\epsilon_{-}(p) = \epsilon_{0}(p) + \mu_B H$, respectively.
Two Cooper pairs with spin parts of internal wave functions,
 $\Delta(+, -)$ and  $ \Delta(-, +)$, and total momenta $p_0 
 \neq 0$ are characterized by different probabilities to exist
 since energy difference $|\epsilon^{1}_{+} - \epsilon^{1}_{-}|
 = p_0 v_F + 2 \mu_B H$ is not equal to energy difference 
$|\epsilon^{2}_{-} - \epsilon^{2}_{+}|
= - p_0 v_F + 2 \mu_B H$  if $p_0 \neq 0$.
[For simplicity, linearized one-dimensional electron spectrum,
$\epsilon (p) = v_F |p|$, is considered].
}
\label{fig2}
\end{figure}


\begin{references}

\bibitem{AAA} A.A. Abrikosov, 
{\it Fundamentals of Theory of Metals} (Elsevier Science
Publisher B.V., Amsterdam, 1988).

\bibitem{PGG} P.G. de Gennes, 
{\it Superconductivity of Metals and Alloys} 
(Westview Press, Advanced Book Program, 
Boulder , 1999).

\bibitem{GorMel}   L.P. Gorkov and T.K. Melik-Barkhudarov, 
Zh. Eksp. Teor. Fiz. \underline{45}, 1493 (1963)
[Sov. Phys. JETP, \underline{18}, 1031 (1964)] . 

\bibitem{JMRB} D. Jerome, A. Mazaud, M. Ribault, and
K. Bechgaard, 
J. Phys. (Paris) Lett. \underline{41}, L-95 (1980). 


\bibitem{IYS} T. Ishiguro, K. Yamaji, and G. Saito, 
{\it Organic Superconductors} 
(Second Edition, Springer-Verlag, Heidelberg, 1998).

\bibitem{SABL} F. Steglich, J. Aarts, C.D. Bredl, W. Leike,
D. Meschede, W. Franz, and H. Schafer,
Phys. Rev. Lett. \underline{43}, 1892 (1979). 

\bibitem{BedMul} J.G. Bednortz and K. Muller,
Z. Rev. B  \underline{64}, 189 (1986). 

\bibitem{Maeno} Y. Maeno, H. Hashimoto, K. Yoshida, 
S. Nishizaki, T. Fujita, J.G. Bednortz and F. Lichtenberg,
Nature  \underline{372}, 532 (1994). 

\bibitem{Nagamatsu}   J. Nagamatsu, N. Nakagawa, 
T. Muranaka, Y. Zenitani, and J. Akimitsu, 
Nature \underline{410}, 63 (2001). 

\bibitem{SigUeda} M. Sigrist and K. Ueda, 
Rev. Mod. Phys.  \underline{63}, 239 (1991). 


\bibitem{MinSam}   V.P. Mineev and K.V. Samokhin, 
{\it Introduction to Unconventional Superconductivity} 
(Gordon and Breach Science Publishers, Australia, 
1999).

\bibitem{Type-III} We do not use term "type-III superconductivity"
in order to distinguish between suggested in the Letter novel
bulk phenomenon and a so-called surface superconductivity,
which is characterized by critical field $H_{c3}(T)$ [1,2].


\bibitem{Clog} A.M. Clogston, 
Phys. Rev. Lett.  \underline{9}, 266 (1962); 
B.S. Chandrasekhar, 
Appl. Phys. Lett.  \underline{1}, 7 (1962). 

\bibitem{A.G.Lebed} Type-IV superconductivity
phenomenon in layered d-wave superconductors
is considered in A.G. Lebed, Phys. Rev. Lett.,
submitted (2005).

\bibitem{Gorkov-1} L.P. Gorkov, 
Zh. Eksp. Teor. Fiz. \underline{34}, 735 (1958) 
[Sov. Phys. JETP \underline{7}, 505 (1958)].

\bibitem{AbrGor} A.A. Abrikosov, L.P. Gorkov,
and I.E. Dzyaloshinskii,
{\it Methods of Quantum Field Theory in
Statistical Physics}
(Dover Publications, New York, 1963). 

\bibitem{Gorkov-3} L.P. Gorkov, 
Zh. Eksp. Teor. Fiz. \underline{37}, 833 (1959) 
[Sov. Phys. JETP \underline{37}, 593 (1960)].

\bibitem{Gorkov-2} L.P. Gorkov, 
Zh. Eksp. Teor. Fiz. \underline{36}, 1918 (1959) 
[Sov. Phys. JETP \underline{36}, 1364 (1959)].

\bibitem{Abrikosov-2} A.A. Abrikosov, 
Zh. Eksp. Teor. Fiz. \underline{32}, 1442 (1957) 
[Sov. Phys. JETP \underline{5}, 1174 (1957)].

\bibitem{SalVol} M.M. Salomaa and G.E. Volovik,
Rev. Mod. Phys. \underline{59}, 533 (1987). 


\bibitem{MatSimNag} S. Matsuo, H. Shimahara,
and K. Nagai,
J. Phys. Soc. Jpn. \underline{63}, 2499 (1994). 


\bibitem{ShimaPhys} H. Shimahara,
Phys. Rev. B \underline{62}, 3524 (2000). 

\bibitem{GorRab} L.P. Gorkov and E.I. Rashba,
Phys. Rev. Lett. \underline{87}, 037004 (2001). 







\end{references}
\end{document}